\documentclass[acmsmall]{acmart}
\AtBeginDocument{%
 }

\acmJournal{CSUR}

\usepackage{listings}
\usepackage{xcolor}
\usepackage{booktabs}
\usepackage{tikz}
\usepackage{multirow}
\usepackage{booktabs}
\usepackage{pgfplots}
\usepackage{comment}
\usepackage{pgf-pie}
\usetikzlibrary{patterns}
\usetikzlibrary{calc}
\pgfplotsset{compat=1.18}
\usetikzlibrary{positioning}

\lstdefinestyle{sqlstyle}{
	language=SQL,
	basicstyle=\ttfamily\small,
	keywordstyle=\color{blue},
	commentstyle=\color{gray},
	stringstyle=\color{teal},
	showstringspaces=false,
	breaklines=true,
	frame=single,
	framerule=0pt,
	backgroundcolor=\color{gray!5},
	tabsize=2
}

\lstdefinestyle{bashstyle}{
	language=bash,
	basicstyle=\ttfamily\small,
	commentstyle=\color{gray},
	showstringspaces=false,
	breaklines=true,
	frame=single,
	framerule=0pt,
	backgroundcolor=\color{gray!5},
	tabsize=2
}

\begin{document}

\title{Bridging Behavioral Biometrics and Source Code Stylometry: A Survey of Programmer Attribution}

\author{Marek Horv\'ath}
\orcid{0009-0005-4649-2308}
\affiliation{%
 \institution{Department of Computers and Informatics, Technical University of Ko\v{s}ice}
 \city{Ko\v{s}ice}
 \country{Slovakia}
}
\email{marek.horvath@tuke.sk}

\author{Em\'ilia Pietrikov\'a}
\orcid{0000-0002-9790-6874}
\affiliation{%
 \institution{Department of Computers and Informatics, Technical University of Ko\v{s}ice}
 \city{Ko\v{s}ice}
 \country{Slovakia}
}
\email{emilia.pietrikova@tuke.sk}

\author{Diomidis Spinellis}
\orcid{0000-0003-4231-1897}
\affiliation{%
 \institution{Department of Management Science and Technology, Athens University of Economics and Business}
 \city{Athens}
 \country{Greece}
}
\email{dds@aueb.gr}

\renewcommand{\shortauthors}{Horvath et al.}

\begin{abstract}
Programmer attribution seeks to identify or verify the author of a source code artifact using stylistic, structural, or behavioural characteristics. This problem has been studied across software engineering, security, and digital forensics, resulting in a growing and methodologically diverse set of publications. This paper presents a systematic mapping study of programmer attribution research focused on source code analysis. From an initial set of 135 candidate publications, 47 studies published between 2012 and 2025 were selected through a structured screening process. The included works are analysed along several dimensions, including authorship tasks, feature categories, learning and modelling approaches, dataset sources, and evaluation practices. Based on this analysis, we derive a taxonomy that relates stylistic and behavioural feature types to commonly used machine learning techniques and provide a descriptive overview of publication trends, benchmarks, programming languages. A content-level analysis highlights the main thematic clusters in the field. The results indicate a strong focus on closed-world authorship attribution using stylometric features and a heavy reliance on a small number of benchmark datasets, while behavioural signals, authorship verification, and reproducibility remain less explored. The study consolidates existing research into a unified framework and outlines methodological gaps that can guide future work. This manuscript is currently under review. The present version is a preprint.
\end{abstract}

\ccsdesc[500]{Software and its engineering~Empirical software validation}
\ccsdesc[300]{Human-centered computing~Biometrics}
\ccsdesc[200]{Computing methodologies~Supervised learning}
\ccsdesc[100]{Security and privacy~Software and application security}

\keywords{Behavioral biometrics, source code stylometry, programmer identification, authorship attribution, authorship verification, code forensics, programming education, machine learning}

\maketitle

\section{Introduction}

Programmer identification refers to the process of determining or verifying the author of a source code fragment based on measurable characteristics of programming style or behavior. Authorship information can be valuable in multiple contexts, including software forensics, educational analytics, plagiarism detection, developer profiling, and recruitment. A range of approaches has been explored for this purpose, from traditional statistical models to modern machine learning methods designed to capture distinctive features of code.

Most existing studies focus primarily on stylistic aspects of source code, such as lexical, syntactic, or structural patterns. Less attention has been given to behavioral factors that can also reflect individual differences in how programmers write, edit, and manage code. Understanding how these stylistic and behavioral dimensions are represented in research is essential for building a complete view of programmer attribution as a field.

The goal of this survey is to systematically review and organize existing work on programmer identification, with attention to the types of features, data collection strategies, datasets, analytical methods, and evaluation procedures used in the literature. The study also considers behavioral aspects when present, aiming to identify how they have been combined or contrasted with stylometric evidence in prior research.

To achieve this, the study follows established guidelines for secondary research in software engineering~\cite{kitchenham2007guidelines}, applying a structured and reproducible procedure for collecting and categorizing relevant studies.

The mapping study is guided by the following research questions:

\begin{itemize}
	\item RQ1: Which behavioral and stylistic features have been applied in programmer identification research, and how are they categorized?
	\item RQ2: What machine learning and statistical techniques are used to model these features for authorship attribution and verification?
	\item RQ3: In what ways have behavioral biometrics been combined with source code stylometry, and what evidence supports their joint effectiveness?
	\item RQ4: What datasets, preprocessing pipelines, and evaluation protocols are commonly used in this field?
\end{itemize}

Together, these questions aim to outline the current state of research on programmer attribution and to establish a consistent framework for analyzing how behavioral and stylistic evidence have been utilized across different studies.

\begin{figure}[htbp]
	\centering
	\begin{tikzpicture}[
		node distance=1.1cm,
		box/.style={
			rectangle,
			draw=black,
			rounded corners=2pt,
			align=center,
			minimum width=4.7cm,
			minimum height=0.9cm,
			font=\bfseries\small
		},
		arrow/.style={->, thick}
		]
		
		\node[box] (time) {Time Range of the Search};
		\node[box, below=1.1cm of time] (keywords) {Search Keywords};
		\node[box, below=1.1cm of keywords] (criteria) {Inclusion and\\Exclusion Criteria};
		
		\node[box, right=3cm of time] (db) {Selected Databases};
		\node[box, right=3cm of keywords] (alex) {Retrieval articles};
		\node[box, right=3cm of criteria] (extract) {Data Extraction and\\Recorded Attributes};
		
		\draw[arrow] (time) -- (db);
		\draw[arrow] (db) -- (keywords);
		\draw[arrow] (keywords) -- (alex);
		\draw[arrow] (alex) -- (criteria);
		\draw[arrow] (criteria) -- (extract);
		
	\end{tikzpicture}
	\caption{Overview of the systematic mapping process}
	\label{fig:methodology_flow}
\end{figure}
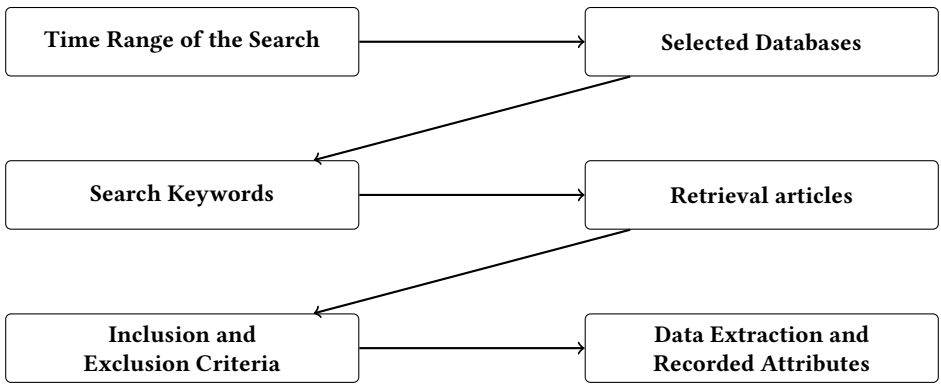

\section{Background}

This section introduces the conceptual foundations required to interpret the remainder of the survey. We first clarify key terms and notions that recur throughout the literature on programmer identification, as their usage is not always consistent across disciplines. We then position the present work with respect to existing survey articles, highlighting differences in scope, focus and analytical depth.

\subsection{Terminology}

The literature on authorship analysis spans multiple research communities, including software engineering, security, forensic analysis and computational linguistics. As a result, core concepts are sometimes defined differently or used interchangeably. The following terminology establishes the meanings adopted in this survey to ensure clarity and consistency in subsequent sections.

\textbf{Authorship attribution.}
Authorship attribution refers to the task of determining who wrote a given piece of text or source code. In the context of software forensics, it represents the process of analyzing programming artifacts to identify their likely author~\cite{tennyson_replicated_2013}. The goal is to associate a code fragment with one programmer among a known set of candidates, typically using features extracted from the code itself. 

\textbf{Authorship verification.}
Authorship verification differs from attribution in that it evaluates whether a given piece of code was written by a claimed author, without necessarily identifying who the actual author is if the claim is false. It is commonly formulated as a binary classification problem, where examples of the suspected author's previous work serve as positive samples and unrelated code constitutes the negative class~\cite{caliskan-islam_-anonymizing_2015}.

\textbf{Static code analysis.}
Static code analysis examines source code without executing it, extracting properties related to structure, syntax, or style. It is used to detect programming errors, quality issues, or stylistic characteristics. This approach enables reproducible analysis of large codebases.

\textbf{Stylometry.}
Stylometry is the quantitative study of individual writing style. In programming, it examines how personal habits in code layout, naming conventions, control structures, and syntactic organization manifest as measurable stylistic patterns~\cite{frankel_machine_2021, gurioli_stylometry_2024}. Stylometric features are commonly grouped as lexical, syntactic, structural, or semantic, and may be extracted through tokenization or analysis of abstract syntax trees (ASTs), which capture the hierarchical structure of source code.

\textbf{Stylochronometry.}
Stylochronometry studies temporal changes in an author's writing style. In source code analysis, it evaluates whether a programmer's stylistic features remain consistent or drift over time, which may affect the reliability of attribution models~\cite{petrik_source_2018, petrik_effect_2021}. The concept is particularly relevant for long-term datasets such as programming competitions or version control histories.

\textbf{Behavioral biometrics.}
Behavioral biometrics focus on identifying individuals through distinctive behavioral patterns, such as typing dynamics, edit sequences, or commit activity. In software engineering, such data capture temporal and interactive aspects of programming that complement static code stylometry~\cite{caliskan-islam_-anonymizing_2015}. 

\textbf{Feature extraction.}
Feature extraction is the process of transforming raw source code into quantitative representations suitable for analysis. In authorship studies, features can be derived from lexical tokens, syntax trees, or style metrics that capture the individual habits of a programmer.

\textbf{Lexical features.}
Lexical features describe surface level elements of source code, including keywords, identifiers and operators, as well as character based statistics. They capture stylistic tendencies through the frequency and ordering of tokens and are often considered relatively language independent.

\textbf{Syntactic features.}
Syntactic features capture how a programmer structures code statements and control flows according to programming language grammar. They are often derived from parse trees or abstract syntax trees and reflect the author's habits in constructing logical expressions or loops.

\textbf{Structural features.}
Structural features represent higher-level organization of code, including class design, function decomposition, nesting depth, or complexity metrics. They offer insights into an author's design preferences beyond syntax or vocabulary.

\textbf{Semantic features.}
Semantic features express meaning-oriented properties of code, such as the use of specific algorithms, data structures, or naming semantics. They attempt to capture the conceptual level of programming choices, complementing syntactic and structural representations.

\textbf{Tokenization.}
Tokenization is the process of segmenting source code into discrete lexical units, such as keywords, identifiers, or operators. It serves as the foundation for extracting lexical and syntactic features and for computing frequency-based representations.

\textbf{n-gram analysis.}
n-gram analysis divides a sequence of tokens or characters into contiguous subsequences of length \(n\). In authorship analysis, it identifies recurring token combinations that indicate consistent stylistic habits. The technique is widely used for both text and source code stylometry.

\textbf{Abstract Syntax Tree (AST).}
An Abstract Syntax Tree is a hierarchical representation of source code structure in which each node corresponds to a language construct. It preserves the syntactic relationships between statements and enables systematic extraction of stylistic and structural features for authorship analysis.

\textbf{Feature representation.}
Feature representation refers to encoding extracted features into numerical vectors that can be processed by statistical or machine learning models. Typical representations include frequency distributions, TF-IDF weighting, or embeddings derived from code structures.

\subsection{Related work}

Authorship analysis has been the subject of several survey and review articles that examine stylometric techniques across different data types and application contexts. This section briefly reviews representative surveys to position the present work within the existing literature. 

Neal et al.~\cite{neal_surveying_2018} present a broad survey of stylometry with a primary focus on natural language rather than source code. Source code is mentioned only marginally, and the survey does not analyse feature families, representations, datasets or evaluation practices that are specific to programmer identification. The work addresses multiple authorship related tasks, including attribution, verification, profiling and style change detection, and organises stylometric features into lexical, syntactic, semantic, structural and domain specific categories. A wide range of statistical, machine learning and neural techniques is reviewed, together with common limitations such as sensitivity to short texts, scalability issues and the absence of standardised benchmarks. While the survey provides useful background on stylometric methods in general, it does not offer a taxonomy or synthesis tailored to source code authorship analysis.

Another selected survey by Alhijawi et~al.~\cite{alhijawi_text-based_2018} reviews authorship identification research from 2007 to 2017 with an emphasis on forensic and security applications. The authors categorise prior work according to text modality, including email communication, online messages, blogs and instant messaging, and treat authorship identification as one task within a broader authorship analysis landscape. Stylometric features are organised into lexical, character level, syntactic and semantic categories, and the dominance of datasets such as the Enron email corpus and PAN competition collections is noted. Modelling techniques are summarised descriptively, with classification based approaches such as support vector machines, k nearest neighbours, Naïve Bayes and decision trees appearing most often. Source code is mentioned only briefly, and there is no structured comparison of code specific features, models or evaluation protocols. Behavioural signals and their integration with stylometric features fall outside the scope of this survey.

In contrast, He et~al.~\cite{he_authorship_2024} provide a broad review of authorship attribution across natural language texts, source code and mixed digital artefacts. Their survey is organised along multiple analytical dimensions, including feature extraction strategies, learning models, datasets, evaluation metrics and recurring challenges, and devotes considerable attention to the rise of deep learning techniques and their impact on attribution accuracy. Issues such as class imbalance, scale and the scarcity of benchmarks are highlighted. Although source code authorship is included, it is treated as one of several domains. Behavioural aspects are acknowledged as open challenges and the integration of programmer habits with code stylometry is left largely unexplored.

Kalgutkar et~al.~\cite{kalgutkar_code_2020} offer a focused survey of code authorship attribution with an emphasis on malware analysis, software forensics and adversarial contexts. The authors differentiate attribution from plagiarism detection and trace the evolution of the field from early software metrics and similarity measures to modern machine learning based approaches on source code and binaries. A central contribution of their survey is a taxonomy of feature types, including lexical, syntactic, semantic, behavioural and application specific features, and an analysis of how each category captures different aspects of programming style. Lexical and syntactic features are identified as the most common due to their simplicity and availability. Behavioural features are interpreted primarily as programmer habits inferred from code rather than explicit interaction data. Learning techniques ranging from k nearest neighbours and support vector machines to neural networks are reviewed, but the survey does not systematically relate models to feature categories, programming languages or evaluation settings. Evaluation practices are summarised at a high level and often assume closed world scenarios and limited author sets. 

Finally, Gray et~al.~\cite{gray_identifying_2024} focus on authorship attribution in malicious binaries, a domain where source code is unavailable and one must rely on compiled artefacts. They distinguish malware authorship attribution from malware family classification and binary similarity analysis, framing authorship attribution as an inherently adversarial and open ended problem. The survey reviews datasets and modelling approaches for malware attribution and discusses limitations such as uncertain ground truth, closed world assumptions and the treatment of threat actor groups as single authors. Source code based programmer identification and the integration of behavioural and stylistic cues in non adversarial settings fall outside its scope.

The reviewed surveys illustrate the diversity of authorship analysis research across natural language, forensic applications, cross-domain settings and malware analysis. At the same time, they reveal a common limitation. Source code is either treated only marginally or examined within narrowly defined application contexts. None of the existing surveys offers a unified and systematic synthesis focused specifically on programmer identification from source code that jointly considers feature types, learning models, datasets and evaluation practices. In particular, behavioural aspects and their integration with traditional code stylometry receive limited attention. These gaps motivate the present study, which provides a dedicated mapping and taxonomy of source code authorship research grounded in a consistent analytical framework.

\section{Review Methodology}

This section outlines the systematic procedure followed to identify, select, and analyze publications relevant to programmer identification. The methodology was designed to ensure transparency, reproducibility, and coverage of the research domain. It describes the temporal scope of the search, the formulation of keyword groups and logical combinations, the choice of digital libraries, and the criteria used for inclusion and exclusion. Subsequent subsections detail the data extraction process and recorded attributes.

\subsection{Time range of the search}

The lower bound of the search period was set to 2012, corresponding to the point when Git-based hosting platforms and mining infrastructures matured into large, publicly accessible ecosystems. Git itself was introduced in 2005, but Spinellis~\cite{spinellis2012git} described GitHub in 2012 as ``\textit{the best known}'' collaborative code hosting service that already promoted large scale cooperation and pull based development workflows. By this time, GitHub had evolved from a simple repository mirror into the dominant platform for open-source collaboration, providing the social and technical infrastructure required for data driven studies of developer behavior. Cosentino~et~al.~\cite{cosentino2016findings} reported that the platform stores more than 35 million projects in 2016 and analyzed research on GitHub mining beginning in 2010, indicating a rapid expansion of available data and research interest during the following years. Hu~et~al.~\cite{hu2016github} similarly referred to millions of open source software projects hosted on GitHub and analyzed event data from 2012-2015, demonstrating that rich behavioral and interaction logs were publicly accessible in this period. Finally, Markovtsev and Long~\cite{markovtsev2018public} documented that by 2017 GitHub contained over 67 million projects and published the \textit{Public Git Archive} dataset of 182\,014 repositories (3.0\,TB), confirming the platform´s transition into an ultra-large corpus suitable for empirical research. Collectively, these observations justify restricting the review window to 2012-2025, which captures the contemporary, data-intensive era of programmer identification and source code stylometry.

\subsection{Selected databases}

The literature search was conducted across three major digital libraries widely used in computer science research: IEEE Xplore, ACM Digital Library, and Scopus. In addition, Google Scholar was used for backward and forward snowballing to identify papers that cite or are cited by the initially selected studies. These sources together ensured coverage of peer-reviewed works relevant to software engineering, code analysis, and authorship attribution. Each database indexes a distinct subset of the research community, reducing the risk of selection bias.

\begin{itemize}
	\item \textit{IEEE Xplore}\footnote{\url{https://ieeexplore.ieee.org/Xplorehelp/overview-of-ieee-xplore}} is a digital library providing access to scientific and technical publications issued by IEEE and its publishing partners. It includes journals, magazines, conference proceedings, books, standards, and selected course materials across electrical engineering, computing, and related technological domains.

	\item \textit{ACM Digital Library}\footnote{\url{https://dl.acm.org/about}} is the publishing and discovery platform of the Association for Computing Machinery. It offers full-text access to ACM journals, conference proceedings, magazines, and newsletters, together with bibliographic indexing and abstracting services through the ACM Guide to Computing Literature.

	\item \textit{Scopus}\footnote{\url{https://www.elsevier.com/products/scopus/content}} is a multidisciplinary abstracting and indexing database covering peer-reviewed serial publications with ISSN identifiers, including journals, conference proceedings, and book series, as well as selected non-serial publications with ISBN identifiers. It aggregates metadata across publishers and disciplines and provides citation-based metrics.
\end{itemize}

\subsection{Search keywords}

The search process was guided by a structured set of keyword groups rather than a single query. This design ensured coverage of terminology relevant to programmer identification, including the research objective, the analyzed object, and the methodological perspective. The keywords were organized into the following groups:

\begin{itemize}
 \item \textit{Group A - Authorship task:} terms describing the goal of identifying or verifying the author of source code.
 \item \textit{Group B - Source code object:} terms ensuring that the search focuses on programming artifacts rather than natural language text.
 \item \textit{Group C - Analytical approach:} terms representing the methodological and analytical aspect of stylometric or behavioral analysis.
\end{itemize}

The complete set of keywords used in the search process is presented in Table~\ref{tab:keywords}.

\begin{table}[htbp]
 \caption{Keyword Groups Used in the Search Process}
 \centering
 \small
 \begin{tabular}{@{}l l l@{}}
 \toprule
 \textbf{Group} & \textbf{Focus} & \textbf{Keywords used in search} \\
 \midrule
 A & Authorship task &
 \begin{tabular}[t]{@{}l@{}}
 ``authorship attribution''\\
 ``authorship verification''\\
 ``programmer identification''\\
 ``developer identification''
 \end{tabular} \\
 \addlinespace[0.6em]
 B & Source code object &
 \begin{tabular}[t]{@{}l@{}}
 ``source code''\\
 ``programming assignments''\\
 ``software repositories''\\
 ``GitHub''
 \end{tabular} \\
 \addlinespace[0.6em]
 C & Analytical approach &
 \begin{tabular}[t]{@{}l@{}}
 ``stylometry''\\
 ``behavioral biometrics''\\
 ``commit behavior''\\
 ``interaction logs''\\
 ``code analysis''
 \end{tabular} \\
 \bottomrule
 \end{tabular}
 \label{tab:keywords}
\end{table}

The final search queries were constructed to ensure that each retrieved publication contained at least one keyword from each of the three groups. 
Formally, the Boolean logic applied during the search can be expressed as:

\[
\begin{array}{c}
(A_1 \ \text{OR}\ A_2 \ \text{OR}\ A_3 \ \text{OR}\ A_4) \\[0.6em]
\text{AND} \\[0.6em]
(B_1 \ \text{OR}\ B_2 \ \text{OR}\ B_3 \ \text{OR}\ B_4) \\[0.6em]
\text{AND} \\[0.6em]
(C_1 \ \text{OR}\ C_2 \ \text{OR}\ C_3 \ \text{OR}\ C_4 \ \text{OR}\ C_5)
\end{array}
\]

where \(A_i\), \(B_i\), and \(C_i\) denote individual keywords from Groups A, B, and C respectively (see Table~\ref{tab:keywords}). This formulation ensures that at least one term from each group appears in the search query, capturing studies that address programmer authorship, source code artifacts, and analytical approaches.

The choice of these particular keywords was grounded in both domain conventions and empirical usage across prior software engineering literature. Terms in Group A reflect the established vocabulary used in authorship analysis and software forensics research, ensuring inclusion of studies that may differ in phrasing but share the same goal of author identification. The expressions in Group B were selected to restrict results strictly to programming-related sources and to avoid retrieving work on natural language stylometry, which often dominates search results when authorship attribution is used alone. Finally, the terms in Group C were derived from two methodological traditions that converge in this review stylometric feature extraction and behavioral or biometric data analysis.

The selection of keywords was also guided by terminology consistently used in the authors' previous empirical studies in related areas. Earlier works on source code stylometry and behavioral biometrics~\cite{horvath2024stylistic,horvath2025programmer}, plagiarism detection based on code decomposition~\cite{horvath2024codeclones}, and personalized learning analytics through static code analysis~\cite{horvath2024aip} employed expressions such as \textit{“source code stylometry,” “behavioral biometrics,” “static code analysis,” and “programmer identification.”} Drawing on this prior experience helped align the search vocabulary with realistic phrasing found in contemporary research and ensured continuity across educational, forensic, and behavioral perspectives relevant to programmer attribution.

The organization of the keyword groups was directly aligned with the review objectives and research questions (RQ1-RQ4) to maintain methodological traceability. Group A corresponds to the task definition (authorship and verification), Group B constrains the analyzed artifact (source code), and Group C targets the analytical approaches relevant to feature extraction and behavioral modeling.

\subsection{Inclusion and exclusion criteria}

To maintain consistency and transparency throughout the selection process, predefined inclusion and exclusion criteria were applied during both abstract and full-text screening. These criteria were derived from the research questions (RQ1-RQ4) and refined through pilot testing of initial search results to ensure they captured studies relevant to programmer identification.

\textbf{Inclusion criteria:}
\begin{itemize}
	\item Published between 2012 and 2025.
	\item Written in English and peer-reviewed (journal article or conference paper).
	\item Focused on authorship attribution, verification, or programmer identification involving source code.
	\item Includes an empirical evaluation with quantitative or comparable metrics.
	\item Provides sufficient methodological detail to support reproducibility and replication.
\end{itemize}

\textbf{Exclusion criteria:}
\begin{itemize}
	\item Studies centered on natural language stylometry (literary or linguistic texts) without source code analysis.
	\item Articles limited to plagiarism detection without explicit author identification.
	\item Papers lacking empirical evidence, datasets, or reproducible methodology.
	\item Non-peer-reviewed or informal sources (e.g., theses, editorials, workshop summaries, or preprints without validation).
	\item Publications for which the full text was unavailable even through institutional access.
\end{itemize}

These criteria were consistently applied during all screening rounds to ensure that the final mapping set included only empirical studies. All bibliographic metadata and coded attributes of the included studies were organized into a structured dataset to support transparency, reproducibility, and potential future meta analyses.

\subsection{Data extraction and recorded attributes}

In addition to automatically imported bibliographic information (such as authors, title, venue, and publication year) collected in Zotero, each reviewed study was manually annotated using a structured extraction form. This ensured consistent documentation of methodological and empirical characteristics across all included papers and facilitated later synthesis of results. 
For every study, the following attributes were recorded:

\begin{itemize}
	\item \textit{Inclusion status:} whether the paper met the defined inclusion criteria (yes / no / unclear).
	\item \textit{Reason for exclusion:} concise justification provided when a paper was removed during screening.
	\item \textit{Relevance to research questions:} indicates which research question(s) (RQ1-RQ4) the study contributes to.
	\item \textit{Authorship task:} specifies whether the paper addresses authorship \textit{attribution}, \textit{verification}, or both.
	\item \textit{Feature type:} classification of analyzed features as \textit{stylistic}, \textit{behavioral}, or \textit{hybrid}.
	\item \textit{Learning model:} type of employed model e.g., classical machine learning (SVM, RF, kNN), neural networks (RNN, CNN, Transformer), or statistical approaches.
	\item \textit{Dataset source:} origin of data such as educational repositories, GitHub, or synthetically generated datasets, including the approximate number of authors and samples when available.
	\item \textit{Evaluation method:} description of performance assessment (e.g., metrics such as accuracy, F1-score, or top-\emph{k} results).
	\item \textit{Key findings:} one or two concise sentences summarizing the most relevant outcomes or methodological contributions.
\end{itemize}

These attributes were selected to enable structured comparison between studies while capturing essential methodological details required to answer the research questions.

\subsection*{Excluded articles}

The exclusion process followed the same structure and was applied consistently across all stages of screening. Table~\ref{tab:exclusion} summarizes the main exclusion categories used during the review.

\begin{table}[htbp]
 \caption{Exclusion categories applied during study screening}
 \centering
 \small
 \begin{tabular}{@{}p{5.2cm} p{7.8cm}@{}}
 \toprule
 \textbf{Criterion} & \textbf{Description} \\
 \midrule
 Not related to authorship or programmer identification &
 Paper addresses software metrics, quality, or defect prediction but not author identification or verification. \\
 
 Not focused on source code &
 Study analyzes natural language, documentation, or social media rather than program code. \\
 
 No behavioral or stylistic analysis &
 Focuses on technical performance, tools, or compilers without modeling style or developer behavior. \\
 
 Not empirical / no experimental evaluation &
 Conceptual or theoretical paper without quantitative validation. \\
 
 Plagiarism detection only &
 Detects code similarity or reuse without linking code to an individual author. \\
 
 Duplicate or extended publication &
 Republished version or journal extension of a previously included study. \\
 
 Full text not accessible &
 Paper unavailable even through institutional or academic access. \\
 
 Non-peer-reviewed or non-scientific source &
 Editorial, workshop summary, thesis, or abstract lacking peer review. \\
 
 Irrelevant application domain &
 Study on behavioral biometrics in general security (e.g., keystroke, gait) unrelated to programming activity. \\
 \bottomrule
 \end{tabular}
 \label{tab:exclusion}
\end{table}

The complete set of extracted study attributes, together with the scripts used for parsing BibTeX records and generating the descriptive statistics reported in this review, is publicly available.\footnote{\url{https://doi.org/10.5281/zenodo.18641274}}

\section{Taxonomy of Source Code Authorship Research}

This section analyzes the body of studies that met the inclusion criteria defined in Section~3.4 and were retained after the screening process. Table~\ref{tab:screening} provides an overview of the article collection and filtering steps applied across the selected databases. Starting from the initial pool of retrieved records, successive screening decisions led to a curated set of studies that serve as the basis for the taxonomy and comparative synthesis presented in this section.

\begin{table}[htbp]
	\caption{Article Selection and Screening Process}
	\centering
	\small
	\begin{tabular}{@{}l r l@{}}
		\toprule
		\textbf{Screening phase} & \textbf{Articles} & \textbf{Description} \\
		\midrule
		
		Initial retrieval (IEEE)   & 51  & Search results from IEEE Xplore \\
		Initial retrieval (ACM)    & 73  & Search results from ACM Digital Library \\
		Initial retrieval (Scopus) & 23  & Search results from Scopus database \\

		\addlinespace[0.6em]

		\textit{Total retrieved}   & \textit{147} & \textit{Combined records before screening} \\
		
		\addlinespace[0.6em]
		
		Duplicates removed & 12 & Overlapping entries across databases \\
		Type filtering     & 32 & Non-journal or non-peer-reviewed items \\
		
		Abstract screening   & 43 & Out of scope based on title and abstract \\
		Relevance filtering  & 13 & Surveys or unrelated topics \\
		
		\addlinespace[0.6em]
		
		\textbf{Final included} & \textbf{47} & \textbf{Studies meeting inclusion criteria} \\
		
		\bottomrule
	\end{tabular}
	\label{tab:screening}
\end{table}

Across all screening stages, a total of 135 records were parsed in detail, of which 47 were included and 88 were excluded based on the defined criteria. 
The distribution of exclusion reasons is summarized in Table~\ref{tab:exclusion_reasons}, providing additional transparency regarding the screening decisions.

\begin{table}[htbp]
	\caption{Distribution of Exclusion Reasons}
	\centering
	\small
	\begin{tabular}{@{}l r@{}}
		\toprule
		\textbf{Exclusion reason} & \textbf{Number of studies} \\
		\midrule
		Proceedings or non-journal venues & 32 \\
		Not related to authorship or programmer identification & 23 \\
		Not focused on source code & 15 \\
		Survey article & 6 \\
		Irrelevant application domain & 6 \\
		Focused only on plagiarism detection without authorship attribution & 2 \\
		Not empirical / no experimental evaluation & 1 \\
		Full text not accessible & 1 \\
		Duplicate or extended publication & 1 \\
		Focused only on plagiarism or code cloning & 1 \\
		\bottomrule
	\end{tabular}
	\label{tab:exclusion_reasons}
\end{table}

The subsequent subsections classify the included studies according to authorship tasks, feature types, and learning and modeling approaches. All retained works contribute to Research Questions RQ1, RQ2, and RQ4, while only a smaller subset explicitly addresses behavioral aspects relevant to RQ3, comprising 13 studies, including works by Wang et al.~\cite{wang_integration_2018}, McKnight et al.~\cite{mcknight_style_2018}, and Bayrami et al.~\cite{bayrami_code_2021}. 

\newpage

\subsection{Authorship Tasks}

\begin{figure}[htbp]
\centering
\begin{tikzpicture}[font=\small]

\definecolor{attr}{RGB}{66,133,244}
\definecolor{veri}{RGB}{219,68,55}
\definecolor{overlap}{RGB}{123,80,164}

\def\sideA{5.5}   
\def\sideV{2.8}   
\coordinate (A) at (0,0);   
\coordinate (V) at (3.2,0);  

\path (A) ++(-\sideA/2,-\sideA/2) coordinate (Abl);
\path (A) ++(\sideA/2,\sideA/2)   coordinate (Atr);

\path (V) ++(-\sideV/2,-\sideV/2) coordinate (Vbl);
\path (V) ++(\sideV/2,\sideV/2)   coordinate (Vtr);

\fill[attr, opacity=0.22] (Abl) rectangle (Atr);
\fill[veri, opacity=0.22] (Vbl) rectangle (Vtr);

\begin{scope}
  \clip (Abl) rectangle (Atr);
  \fill[overlap, opacity=0.30] (Vbl) rectangle (Vtr);
\end{scope}

\draw[draw=attr, line width=0.7pt] (Abl) rectangle (Atr);
\draw[draw=veri, line width=0.7pt] (Vbl) rectangle (Vtr);

\node at ($(A)+(-1.8,2.55)$) {\textbf{Attribution}};
\node at ($(V)+(0.6,1.75)$) {\textbf{Verification}};

\node at ($(A)+(-1.7,0)$) {\Large 41};
\node at (2.05,0) {\Large 4};
\node at ($(V)+(0.0,0)$) {\Large 2};

\node[anchor=north west, align=left, text width=5.2cm] at (6.0,2.75) {%
\textbf{Legend}\\[4pt]
\raisebox{0.2ex}{\tikz{\draw[draw=attr, fill=attr, opacity=0.22] (0,0) rectangle (0.35,0.25);}}
\hspace{0.6em} Attribution (\textit{n}=41)\\[6pt]
\raisebox{0.2ex}{\tikz{\draw[draw=overlap, fill=overlap, opacity=0.30] (0,0) rectangle (0.35,0.25);}}
\hspace{0.6em} Attribution + Verification (\textit{n}=4)\\[-1pt]
{\scriptsize \cite{caliskan-islam_-anonymizing_2015, hozhabrierdi_zero-shot_2020, idialu_whodunit_2024, white_deep_2021}}\\[6pt]
\raisebox{0.2ex}{\tikz{\draw[draw=veri, fill=veri, opacity=0.22] (0,0) rectangle (0.35,0.25);}}
\hspace{0.6em} Verification (\textit{n}=2)\\[-1pt]
{\scriptsize \cite{ou_veribin_2024, romanov_integrated_2025}}%
};

\end{tikzpicture}
\caption{Distribution of authorship tasks addressed in the included studies.}
\label{fig:authorship_tasks_distribution_squares}
\end{figure}
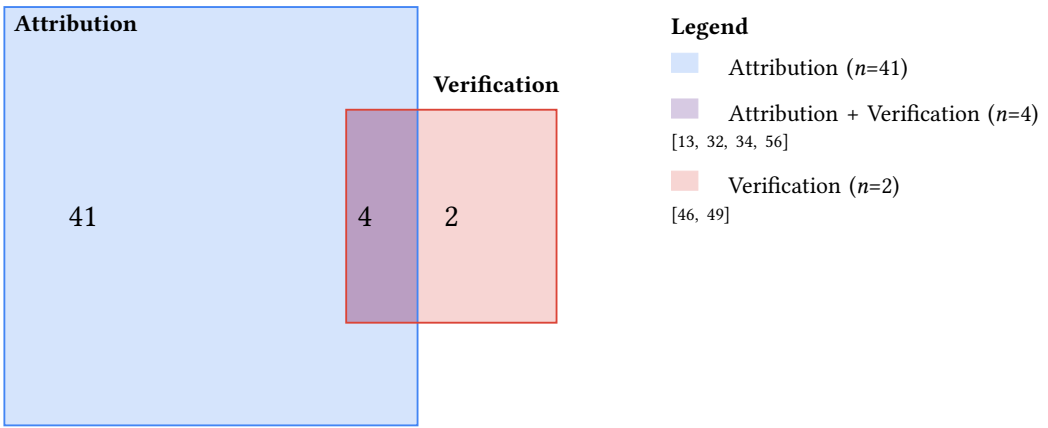

Authorship tasks in programmer identification fall into two principal categories:
\begin{itemize}
	\item \textit{Authorship attribution}, which assigns an unknown code fragment to one programmer from a known set.
	\item \textit{Authorship verification}, which assesses whether a given code fragment matches a claimed author.
\end{itemize}
A small number of studies address both tasks by first verifying a claimed author and subsequently performing attribution when verification fails. The proportions observed in our survey are illustrated in Figure~\ref{fig:authorship_tasks_distribution_squares}, where attribution clearly dominates. Authorship verification on its own is rarely investigated, indicating that this task remains comparatively underexplored.

The overwhelming focus on attribution can be attributed to the relative availability of labelled datasets and the ease of evaluating classification accuracy. Verification, by contrast, requires personalised models or template matching and is harder to scale. Nonetheless, verification plays a key role in forensic settings where the question is not \textit{who wrote this?} but \textit{did this person write this?}. As a result, the limited number of verification studies points to a clear opportunity for future research.

\subsection{Feature Types}

Existing approaches to programmer identification extract information from source code in two broad ways
\begin{itemize}
\item \textit{Stylistic features (41 studies)} capture lexical, syntactic and structural patterns that reflect a programmer's coding habits; examples include token frequencies, abstract syntax tree shapes and control-flow structures. These features form the basis of most studies, reflecting the maturity of stylometry.
\item \textit{Hybrid approaches (6 studies)} enrich stylometry with behavioural signals such as keystroke dynamics, editing sequences or commit histories, aiming to integrate how code is produced with how it looks. Such methods appear infrequently but illustrate a growing interest in behavioural biometrics \cite{romanov_integrated_2025, zafar_language_2020, adam_learning_2023, ou_veribin_2024, abuhamad_large-scale_2021, wang_integration_2018}.
\end{itemize}

The dominance of purely stylistic features in our corpus is unsurprising since stylometry has a long history and provides descriptors that generalise across projects and languages. Lexical and syntactic patterns are easy to compute and require only static analysis of source code. By contrast, behavioural signals rely on instrumented environments or access to development histories, which limits the number of available datasets. Consequently, only six included papers experiment with hybrid representations, typically augmenting stylometric features with keystroke or commit timing information. These hybrid approaches hint at a promising future direction where behavioural biometrics complement traditional code stylometry to capture temporal aspects of coding style. However, their rarity underscores the challenges of collecting and sharin

\subsection{Learning and Modeling Approaches}

Research on programmer attribution employs a wide range of modelling strategies that differ in complexity, data requirements, and interpretability. Based on the reviewed studies, these approaches can be grouped into three broad directions
\begin{itemize}
\item \textbf{Classical machine-learning methods}. These include algorithms such as random forests, support-vector machines, nearest-neighbour classifiers, Naive Bayes, decision trees, and logistic regression. Classical models remain widely used because they are comparatively easy to interpret, perform well with limited training data, and can be implemented with modest computational overhead. In our survey, they account for roughly two-thirds of all modelling instances. Their continued use reflects the strong performance of carefully designed feature representations and the preference for models whose decision mechanisms can be more readily examined in forensic and analytical settings.
\item \textbf{Neural network-based models}. This family includes deep learning architectures such as feed-forward networks, recurrent neural networks (RNNs) \cite{liu_practical_2021}, long short-term memory networks (LSTMs) \cite{chopra_authattlyzer_2022}, convolutional neural networks (CNNs), graph neural networks (GNNs), and transformer models \cite{dipongkor_can_2024}. These approaches are designed to learn hierarchical and semantic patterns directly from raw or lightly processed code representations. The increasing number of neural network-based studies mirrors broader developments in software engineering research and represents approximately one third of the modelling approaches observed in our corpus. At the same time, these methods typically require substantial computational resources and large annotated datasets, which can constrain their applicability in practice.
\item \textbf{Other and specialised techniques}. In addition to the dominant paradigms, several studies explore statistical methods such as SCAP profiles \cite{tennyson_choosing_2014} and Latent Dirichlet Allocation (LDA) \cite{alrabaee_cpa_2020, abazari_dataset_2023}, unsupervised clustering techniques \cite{layton_authorship_2014, layton_unsupervised_2012}, and combinations of multiple learners. Although these approaches appear less frequently, they illustrate the methodological diversity of the field and ongoing efforts to address specific data characteristics or analytical goals. For instance, statistical profile-based methods \cite{yokomori_empirical_2023} are often applied when training data are limited, while unsupervised techniques enable exploratory analysis without relying on labelled authorship information.
\end{itemize}

Taken together, the observed distribution of modelling approaches indicates that the field does not converge on a single dominant technique. Classical machine-learning methods continue to serve as reliable baselines, whereas neural models provide alternative ways to capture more complex stylistic patterns at the cost of increased resource demands. The presence of less common modelling strategies suggests that programmer attribution research remains exploratory, with ongoing experimentation driven by dataset properties and application-specific constraints.

\section{Overview of the Reviewed Literature}

This section provides a descriptive overview of the reviewed literature, focusing on publication trends, datasets, modeling practices, and community characteristics. 
Rather than evaluating individual studies, the analysis aggregates observable patterns across the included papers to contextualize the taxonomy presented in the previous section and to highlight dominant practices and emerging directions in the field.

\subsection{Publication Timeline and Venues}

\begin{figure}[htbp]
\centering
\begin{tikzpicture}
\begin{axis}[
 width=0.8\linewidth,
 ybar,
 symbolic x coords={2012,2013,2014,2015,2017,2018,2019,2020,2021,2022,2023,2024,2025},
 xtick=data,
 xlabel={Year},
 ylabel={Number of papers},
 ymin=0,
 ymajorgrids=true,
 bar width=8pt,
 tick label style={rotate=45,anchor=east},
 ]
\addplot coordinates {(2012,2) (2013,1) (2014,2) (2015,1) (2017,1) (2018,5) (2019,3) (2020,6) (2021,7) (2022,4) (2023,4) (2024,6) (2025,5)};
\end{axis}
\end{tikzpicture}
\caption{Annual number of included papers. }
\label{fig:publications_per_year}
\end{figure}
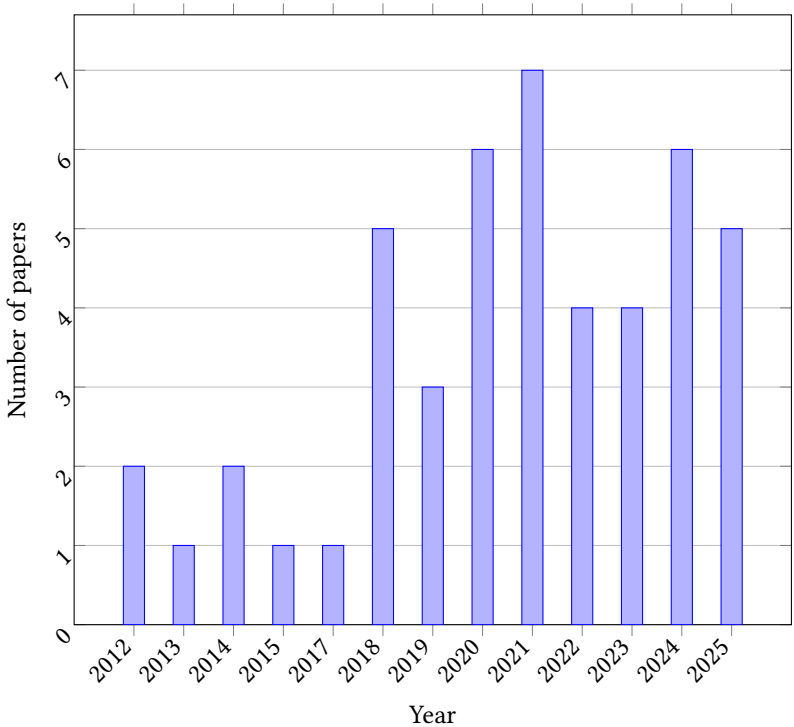

The publication record of programmer attribution research reveals how interest in the field has evolved and where it has been disseminated. Figure~\ref{fig:publications_per_year} summarises the number of included papers published each year between 2012 and~2025. Although early work appeared sporadically, with one or two papers per year until 2017, there is a clear uptick beginning in 2018. The years~2020 and~2021 show the highest productivity, each with six or seven papers, reflecting interest that coincides with the wider adoption of machine learning and neural techniques in software engineering. The persistence of four to six papers per year through~2025 indicates that programmer attribution remains an active area of research rather than a trend.

Venue selection provides additional context about where this work is published. Conference venues dominate with 32 papers, while journals account for 15. This imbalance suggests that the field is largely shaped by conference contributions, which often emphasize timely methodological developments and exploratory evaluations. Journal articles are less frequent but typically offer more extensive analyses, larger datasets, and more comprehensive experimental validation. Together, this distribution reflects both the evolving nature of the field and the continued importance of dissemination alongside more mature journal publications.

\subsection{Datasets and Programming Languages}

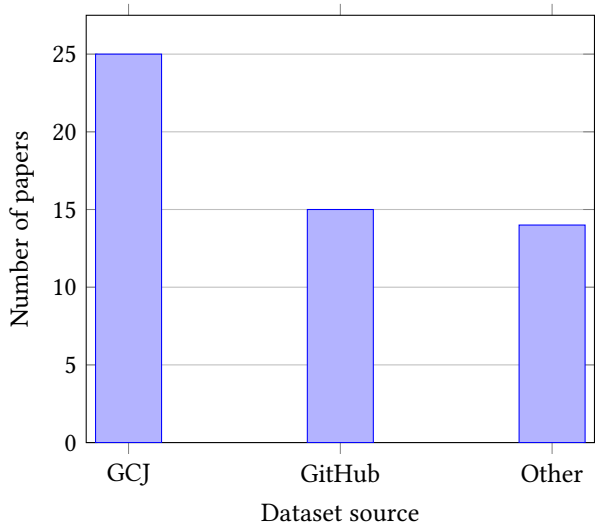
\begin{figure}[htbp]
\centering
\begin{tikzpicture}
\begin{axis}[
 width=0.6\linewidth,
 ybar,
 symbolic x coords={GCJ,GitHub,Other},
 xtick=data,
 xlabel={Dataset source},
 ylabel={Number of papers},
 ymin=0,
 ymajorgrids=true,
 bar width=25pt,
 ]
\addplot coordinates {(GCJ,25) (GitHub,15) (Other,14)};
\end{axis}
\end{tikzpicture}
\caption{Dataset source categories for the included studies.}
\label{fig:dataset_distribution_bar}
\end{figure}

Datasets and programming languages underpin the reproducibility and generality of programmer attribution research. Figure~\ref{fig:dataset_distribution_bar} groups the included studies by their primary data source. The \textit{Google Code Jam} (GCJ) dataset remains the most popular benchmark, used by 25 papers. Its dominance stems from its size, public availability and the well-defined programming tasks it contains, which facilitate consistent comparisons across studies. GitHub repositories constitute the second largest category (15 papers). Studies using GitHub \cite{chatzicharalampous_author_2012} typically harvest code from open-source projects to capture more realistic development styles. The remaining 14 papers rely on synthetic datasets. Diversity suggests that researchers often construct custom datasets to suit specific experimental needs or to explore particular domains such as malware.

Language support is another indicator of the field's breadth. In our corpus, 28 papers do not specify a language, reflecting the widespread use of language-agnostic features or multi-language datasets. Among those that do, C++ \cite{abuhamad_code_2019, alrabaee_cpa_2020, bayrami_code_2021, hajihosseinkhani_authattlyzer-v2_2025} and Java \cite{al-ahmad_meta-heuristic_2023, tennyson_replicated_2013, wang_a3ident_2020, yokomori_empirical_2023} each appear in four papers, while Python is the focus of two studies \cite{balla_code_2024, hozhabrierdi_python_2018}. Nine studies explicitly work with mixed-language datasets. 

\begin{table}[htbp]
\centering
\caption{Occurrence of programming languages and code representations in the included studies. }
\label{tab:language_representation_stats}
\begin{tabular}{lcc}
\toprule
Category & Subcategory & Number of papers \\
\midrule
\multirow{5}{*}{Programming languages} & Unspecified & 28 \\
& Mixed & 9 \\
& C++ & 4 \\
& Java & 4 \\
& Python & 2 \\
\midrule
\multirow{5}{*}{Code representations} & AST & 6 \\
& Embeddings & 3 \\
& Tokens & 2 \\
& CST & 1 \\
& PDG & 1 \\
\bottomrule
\end{tabular}
\end{table}

Code representation choices influence feature extraction and shape how stylistic and structural information is captured from source code. Among the reviewed studies, the abstract syntax tree (AST) is the most commonly used intermediate representation, appearing in six papers \cite{alsulami_source_2017, balla_code_2024, cernansky_comparing_2025, petrik_effect_2021, tennyson_replicated_2013, zafar_language_2020}. Embedding-based representations are employed in three studies \cite{abuhamad_code_2019, white_deep_2021, zafar_language_2020}, while token-based approaches appear in two papers \cite{cernansky_comparing_2025, al-ahmad_meta-heuristic_2023}. The concrete syntax tree (CST) \cite{balla_code_2024} and program dependence graph (PDG) \cite{ullah_source_2019} are each used in a single study. Table~\ref{tab:language_representation_stats} summarises the observed occurrences of both programming languages and code representations. The prevalence of unspecified or mixed programming languages indicates that many approaches aim for language-independent analysis, while the frequent use of ASTs reflects a preference for structured representations that preserve syntactic hierarchy and can be extracted consistently across different languages.

\subsection{Models, Representations, and Experimental Setups}

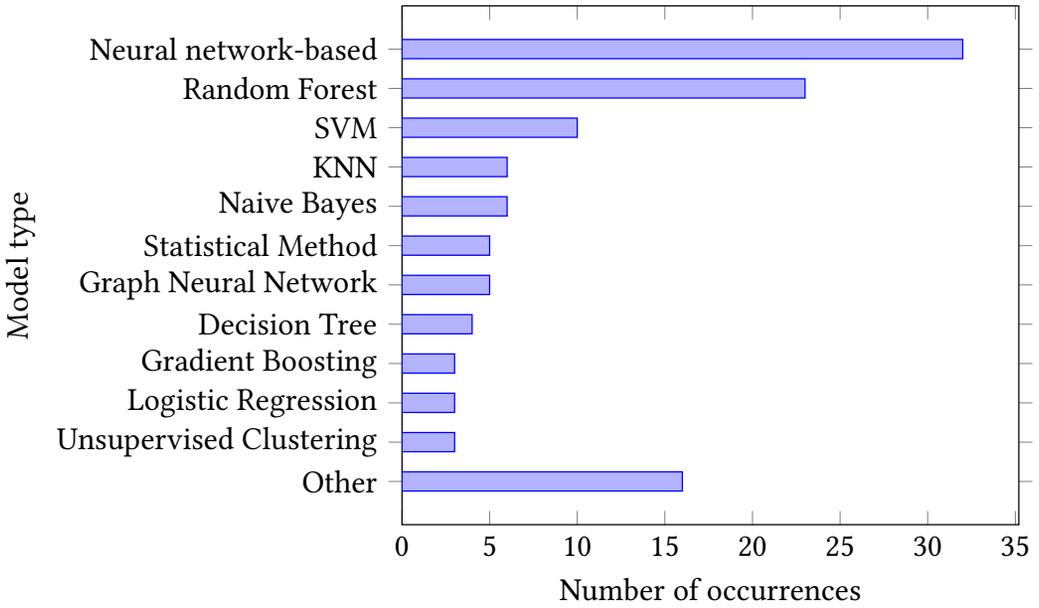
\begin{figure}[htbp]
\centering
\resizebox{\linewidth}{!}{%
\begin{tikzpicture}
\begin{axis}[
  xbar,
  symbolic y coords={Other, Unsupervised Clustering, Logistic Regression, Gradient Boosting, Decision Tree, Graph Neural Network, Statistical Method, Naive Bayes, KNN, SVM, Random Forest, Neural network-based},
  ytick=data,
  xlabel={Number of occurrences},
  ylabel={Model type},
  xmin=0,
  enlarge y limits=0.1,
  bar width=6pt,
]
\addplot coordinates {(16,Other) (3,Unsupervised Clustering) (3,Logistic Regression) (3,Gradient Boosting) (4,Decision Tree) (5,Graph Neural Network) (5,Statistical Method) (6,Naive Bayes) (6,KNN) (10,SVM) (23,Random Forest) (32,Neural network-based)};
\end{axis}
\end{tikzpicture}%
}
\caption{Distribution of learning models across the included studies.}
\label{fig:learning_models_bar}
\end{figure}

Whereas the previous subsection groups modelling strategies at a conceptual level, this subsection examines their concrete use in experimental studies. Figure~\ref{fig:learning_models_bar} reports the frequency of individual learning models across the reviewed literature. Neural network-based approaches appear in 32 studies, while classical algorithms remain prominent at the method level, with random forests used in 23 studies, support-vector machines in 10 studies, and k-nearest neighbours in 6 studies. Naive Bayes classifiers and statistical models occur less frequently but continue to be part of experimental practice. A further 16 studies rely on methods grouped under \textit{Other}, including fine-tuned language models, Bayesian optimisation \cite{jacobsen_optimization_2021}, LDA, IR techniques, and explicit explainability components. The distribution indicates that most experimental setups are built around a small set of established algorithms, while alternative techniques tend to appear in more specialised or exploratory contexts.

\begin{figure}[htbp]
\centering
\begin{tikzpicture}
\begin{axis}[
	width=0.85\linewidth,
	xbar,
	symbolic y coords={Other,AUC,Recall,Precision,F1-score,Accuracy,Cross-validation},
	ytick=data,
	xlabel={Number of studies},
	xmin=0,
	enlarge y limits=0.15,
	bar width=8pt,
]
\addplot coordinates {
	(13,Other)
	(3,AUC)
	(11,Recall)
	(11,Precision)
	(13,F1-score)
	(32,Accuracy)
	(44,Cross-validation)
};
\end{axis}
\end{tikzpicture}
\caption{Frequency of evaluation practices reported in the included studies. A single study may employ multiple evaluation measures.}
\label{fig:evaluation_methods_bar}
\end{figure}
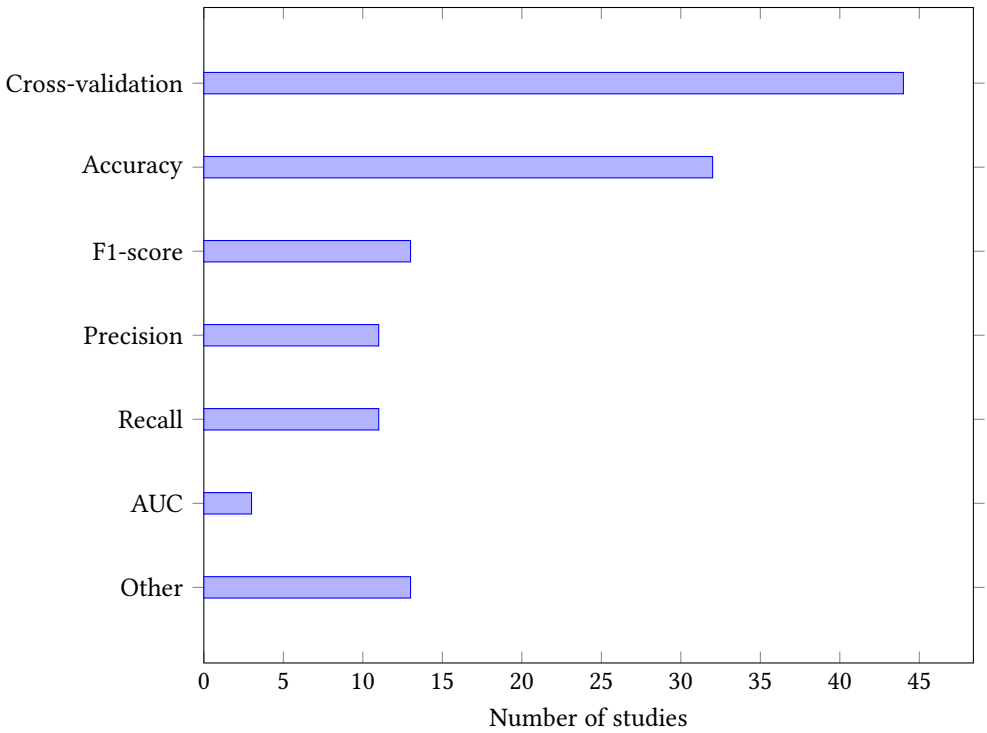

The ways in which studies assess model performance provide additional insight into prevailing experimental practices. Figure~\ref{fig:evaluation_methods_bar} summarises the evaluation measures reported in the reviewed literature. Most studies rely on resampling-based evaluation schemes, while performance is primarily quantified using standard classification metrics such as accuracy, F1-score, precision, and recall. A smaller subset of papers reports alternative measures, including hold-out evaluation, confidence drop, misidentification rate, success rate, overfitting indicators, or training time. This distribution suggests a relatively narrow set of commonly adopted evaluation practices, with limited attention given to complementary criteria such as uncertainty estimation or behaviour under adversarial conditions.

\subsection{Systems and Explainability}

In addition to modelling approaches, several studies describe the development of a concrete system, tool, or framework that implements programmer attribution methods. Nineteen reviewed papers report such systems \cite{aydogan_android_2024}, ranging from small-scale prototypes to more complete pipelines supporting data extraction, feature computation, and classification. However, only a minority of these systems are made publicly available, and most are presented primarily as research artefacts rather than reusable software, which limits their use for replication or direct comparison.

Explainability is addressed even less frequently. Only a small subset of studies integrates explicit explanation techniques, most notably SHAP \cite{hajihosseinkhani_authattlyzer-v2_2025} and LIME \cite{murenin_explaining_2020}, to indicate which features contribute most to attribution decisions. The limited presence of such mechanisms suggests that interpretability is often treated as secondary to predictive performance, even though programmer attribution is frequently discussed in security or forensic settings. Broader adoption of accessible system implementations together with explanatory components would support more transparent and comparable attribution research.

\subsection{Content-Level Descriptors}

\begin{figure}[htbp]
\centering
\begin{tikzpicture}
\begin{axis}[
 width=0.85\linewidth,
 ybar,
 symbolic x coords={feature extraction,machine learning,authorship attribution,codes,source coding,source code,deep learning,training,java,syntactics},
 xtick=data,
 xticklabel style={rotate=45,anchor=east},
 ylabel={Number of papers},
 xlabel={Keyword},
 ymin=0,
 ymajorgrids=true,
 bar width=10pt,
 ]
\addplot coordinates {(feature extraction,14) (machine learning,14) (authorship attribution,13) (codes,12) (source coding,9) (source code,9) (deep learning,9) (training,8) (java,8) (syntactics,7)};
\end{axis}
\end{tikzpicture}
\caption{Top keywords across the included papers.}
\label{fig:top_keywords}
\end{figure}
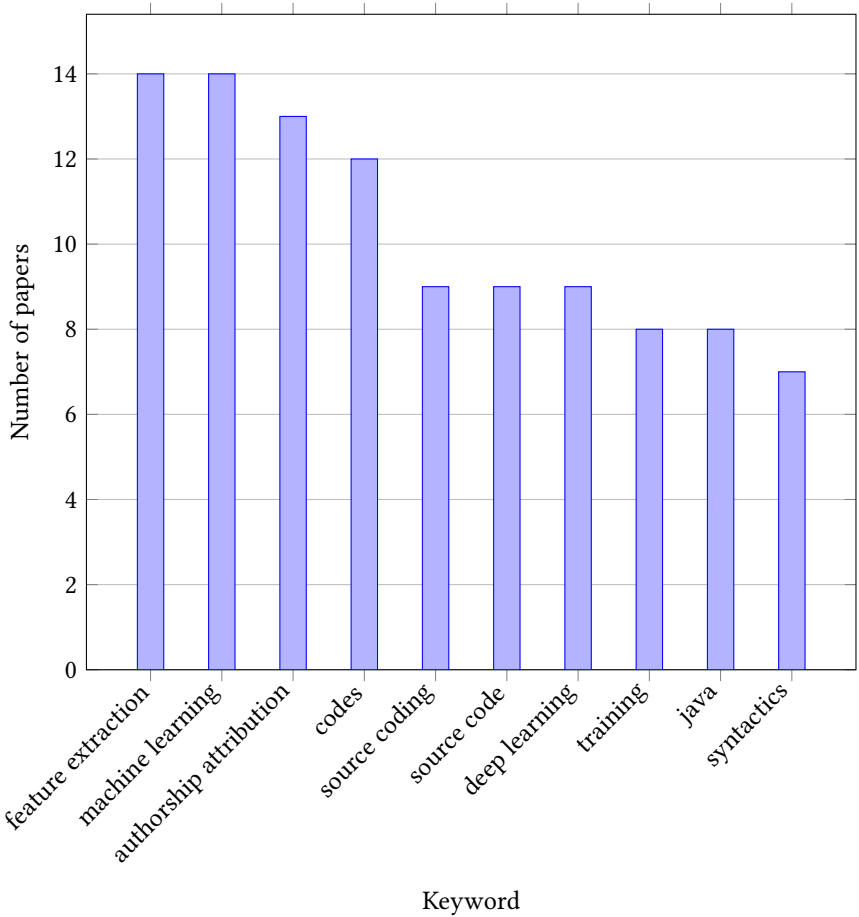

Understanding the language used in publications helps to characterise the thematic focus of programmer attribution research. Figure~\ref{fig:top_keywords} summarises the most frequent keywords across the included papers and highlights the dominant themes addressed in the literature. The prominence of terms such as \textit{feature extraction} and \textit{machine learning} reflects a strong focus on designing informative representations and applying data-driven modelling techniques. The frequent appearance of \textit{authorship attribution} indicates that many studies concentrate on this task while framing their contributions through methodological or technical perspectives. Keywords related to software artefacts, including \textit{codes}, \textit{source coding}, and \textit{source code}, emphasise the close connection to program analysis, whereas the presence of \textit{deep learning} points to the growing adoption of neural approaches. Additional terms such as \textit{training}, \textit{Java}, and \textit{syntactics} suggest attention to model optimisation, language-specific investigations, and syntactic feature design.

\begin{figure}[htbp]
\centering
\includegraphics[width=0.85\linewidth]{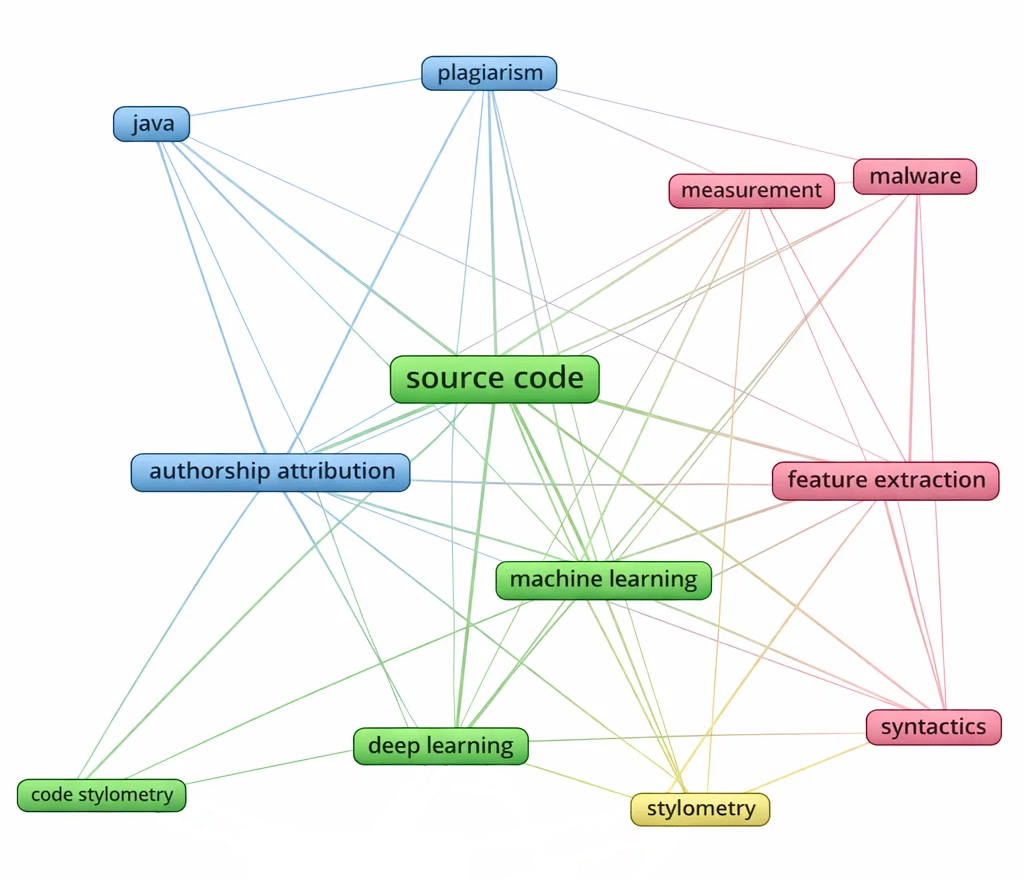}
\caption{Keyword co-occurrence network, illustrating major thematic clusters.}
\label{fig:vos_keywords}
\end{figure}

Beyond individual keyword frequencies, thematic relationships between keywords were analysed using a co-occurrence network constructed with the VOSviewer tool. Figure~\ref{fig:vos_keywords} visualises clusters of frequently co-occurring terms, revealing three closely related thematic areas centred around authorship attribution tasks, machine-learning-based modelling approaches, and feature extraction with application-specific contexts. The central position of \textit{source code} highlights its unifying role across these research directions.

The relationship between titles and abstracts provides additional insight into how authors predominately communicate their contributions. We quantified this relationship by counting the number of shared non-stopwords between each title and its corresponding abstract. The observed overlap ranges from two to ten words, with seven shared terms being the most frequent case, followed by four shared terms. Only a single paper exhibits an overlap of two words, while a smaller number of papers contain eight or more shared terms. The average overlap of 5.85 words indicates that titles generally reflect key terminology introduced in the abstract without directly repeating it, suggesting a balance between descriptive clarity and conciseness.

\subsection{Authorship and Collaboration Characteristics}

Authorship and collaboration patterns provide insight into how research on programmer attribution is organised. Across the reviewed corpus, 147 unique authors contribute to the 47 included studies, indicating a broadly distributed research community. Single-author papers are rare (2 studies), while two-author and three-author teams appear in 11 and 12 cases respectively. The most common configuration consists of teams with four or more authors (22 studies). This distribution suggests that programmer attribution research is typically conducted by small to medium-sized teams, reflecting the need to combine expertise in data collection, feature engineering, modelling, and, in some cases, cross-institutional collaboration.

Recurring contributors offer a complementary perspective on the structure of the research community. Five authors Mohammed Abuhamad \cite{abuhamad_shield_2025, abuhamad_code_2019}, David Mohaisen \cite{choi_attributing_2025}, Richard Harang \cite{dauber_git_2018}, Rachel Greenstadt, and Natalia Stakhanova \cite{matyukhina_adversarial_2019, gonzalez_authorship_2018} each appear in three papers. Although these authors are associated with multiple studies, the overall distribution remains dispersed, with most contributors appearing only once. This pattern suggests that a small number of active groups coexist with a wide range of occasional contributors, supporting both continuity and methodological diversity within the field.

\section{Discussion}

Our analysis shows a reliance on stylistic descriptors such as lexical frequencies, syntactic signatures and structural patterns. These features are well suited to capturing habitual coding practices and can be obtained from source code without special instrumentation. Only a small fraction of studies venture beyond stylometry to incorporate behavioural measurements like keystroke dynamics or edit timing. While collecting behavioural data introduces practical challenges, its potential to capture how code is produced rather than merely how it appears suggests a valuable direction for future work. Another dimension concerns the intermediate representations used during feature extraction. Abstract syntax trees are predominant, followed by token sequences and learned embeddings. More exotic structures, including program dependence graphs and concrete syntax trees, are rarely used even though they could encode deeper semantic cues.

Most empirical evaluations rely on a small number of benchmark datasets. The Google Code Jam corpus offers a controlled set of contest solutions and is used by more than half of the included studies. Its appeal lies in public availability and uniform problem statements, although this homogeneity may limit how well results transfer to real software projects. GitHub repositories represent the second most common data source, providing more diverse and realistic code but introducing challenges such as multiple contributors per project and inconsistent coding conventions. Other studies rely on datasets tailored to specific contexts, including malware analysis, educational settings, or mobile applications. Many papers either do not specify a single programming language or explicitly work with mixed-language, indicating an emphasis on language-agnostic approaches or heterogeneous data collections. At the same time, all of the commonly used datasets predominantly reflect the practices of experienced programmers, such as competition participants or open-source contributors. This leaves student generated code largely underrepresented, motivating our planned focus on novice programmers and educational datasets in future work, particularly in combination with behavioural data.

The reviewed literature employs a wide spectrum of learning algorithms. Classical statistical and machine learning techniques, including random forests, support vector machines, nearest neighbour classifiers, naïve Bayes and logistic regression, remain popular because they are easy to interpret and perform well when combined with carefully engineered features. At the other end of the spectrum, neural methods such as feed forward networks, recurrent architectures, convolutional architectures and graph neural networks are gaining traction. These models often rely on pretrained encoders trained on large code corpora and excel at capturing hierarchical and semantic patterns directly from raw or lightly processed code. The coexistence of these paradigms underscores that there is no single optimal approach. Choice of algorithm should reflect the complexity of the feature set, the size and diversity of the dataset, and the desired balance between accuracy and interpretability. We also observe exploratory work on statistical profiling, clustering and optimisation techniques. Although they appear less frequently, these methods enrich the methodological toolkit and may be particularly suited to specific domains such as malware analysis or binary code.

In terms of task formulations, authorship attribution, which identifies the most likely author among a set of candidates, constitutes the focus of most works. Authorship verification, which checks whether a claimed author wrote a given piece of code, appears in only two studies, and combined approaches that perform verification before attribution are similarly rare. Yet verification is critical in contexts such as code forensics, academic integrity and malware investigation. The phenomenon of collaboratively written code is almost completely absent from experimental setups, even though modern software development often involves multiple contributors. Research that can attribute fragments to multiple programmers or quantify individual influence within a shared codebase remains largely unexplored.

\subsection{Answers to the Research Questions}

This subsection brings together the main findings of the review by directly addressing the research questions defined earlier. Rather than restating individual results, it highlights recurring patterns and shared observations across the analysed studies, providing a synthesis of the current state of the field.

\textbf{RQ1: Features and their categories.} The reviewed studies indicate that programmer attribution relies mainly on stylistic features extracted from source code, which capture consistent patterns in how programmers express solutions rather than the functional behaviour of programs. Lexical features describe surface characteristics such as token frequencies, character or word $n$-grams, comment usage, and identifier naming. Syntactic features focus on grammatical structure, including parse tree shapes, nesting depth, and the organisation of conditional and loop constructs. Structural features reflect higher-level properties of program organisation, for example function length, number of parameters, use of recursion, and branching patterns. A smaller subset of studies combines these static descriptors with behavioural information. Such hybrid feature sets augment code properties with data related to the act of programming, including keystroke timing, editing actions, or commit histories, and only rarely with runtime observations. These approaches attempt to capture how code is produced and modified, not only its final representation. Behavioural features remain uncommon because their collection requires specialised environments and raises privacy concerns. Consequently, most existing work is based on stylometric features alone, showing that behavioural information is still only marginally incorporated into programmer attribution research.

\textbf{RQ2: Modelling and statistical techniques.} The reviewed studies apply a wide spectrum of modelling techniques that range from classical machine learning to more recent neural approaches. Tree based ensembles, support vector machines and nearest neighbour classifiers are used most frequently and form the methodological core of the field, largely because they integrate well with handcrafted stylometric features and remain computationally efficient. Neural models, including recurrent, convolutional, graph based and transformer architectures, appear less uniformly but reflect a gradual shift toward representation learning from structured code artefacts. In addition, several works rely on probabilistic profiling and clustering methods, mainly in exploratory or data constrained settings. Rather than converging on a single dominant paradigm, the literature reflects parallel use of multiple modelling families, each selected according to dataset size, feature design and experimental constraints.

\textbf{RQ3: Integration of behavioural biometrics.} Only a limited subset of studies integrates behavioural signals with static code features. These works demonstrate that timing information derived from typing, editing or version control activity can complement stylometric descriptors, particularly when distinguishing between authors with similar coding styles or when working with short code fragments. At the same time, the overall body of evidence remains small, primarily due to the difficulty of collecting behavioural data and the lack of shared benchmarks. As a result, behavioural information is currently treated as an extension rather than a standard component of programmer attribution pipelines.

\textbf{RQ4: Datasets, pipelines and evaluation protocols.} Empirical evaluations are dominated by a small number of publicly available datasets, with Google Code Jam and GitHub based datasets accounting for most experiments. Typical pipelines follow a similar structure, consisting of code preprocessing, feature extraction or representation learning, and supervised classification under author disjoint data splits. Evaluation is most often based on standard classification metrics, with accuracy and F1 score reported most consistently. While this relative uniformity simplifies comparison, it also highlights limited dataset diversity and a lack of variation in evaluation design. The reviewed studies rarely explore alternative validation strategies or stress test models outside their original data context, which constrains conclusions about general applicability.

\subsection{Limitations of Current Approaches and Future Work}

Based on the findings of this review, several concrete steps can be taken to address current limitations in programmer attribution research and to improve its relevance and reliability in practice.

\begin{itemize}
 \item \textbf{Diversify dataset sources.} Current evaluations rely heavily on competition and open-source datasets that reflect the work of experienced programmers. Future datasets should cover a wider range of contexts, including educational assignments and should span multiple programming languages. 
 \item \textbf{Extend feature sets beyond stylometry.} Lexical and syntactic features are well understood and widely used, but they capture only part of programmer behaviour. Temporal information, repository context and semantic properties derived from program analysis offer complementary signals that remain underexplored. Studying how these feature types interact may reveal patterns that are not visible through static stylometry alone.
 \item \textbf{Improve evaluation practice and reproducibility.} Experimental setups should clearly describe how training and test data are constructed and how authors are separated between them. Evaluation should not rely solely on accuracy but also report metrics that reflect class imbalance and uncertainty in attribution outcomes. Publishing preprocessing scripts, feature extraction pipelines and trained models would support reproducibility and enable more transparent comparison across studies.
 \item \textbf{Address verification and collaborative authorship.} Most existing work focuses on closed-set attribution, while verification and multi-author scenarios remain largely neglected. Developing methods for authorship verification, anomaly detection and attribution within collaboratively written code would better reflect real-world development practices.
 \item \textbf{Revisit modelling choices as representations evolve.} Pretrained code representations and incremental learning techniques open new possibilities for attribution with limited labelled data. Their use should be evaluated carefully, with attention to stability, data requirements and comparability with established methods.
\end{itemize}

\section{Conclusion}

This survey maps programmer attribution research by examining 135 publications and retaining 47 that satisfied the defined inclusion criteria. The reviewed work spans the period from 2012 to 2025 and each study was analysed across several dimensions, including task formulation, feature type, modelling approach, dataset source, evaluation practice and publication venue. On this basis, the survey organises the literature into a coherent taxonomy and outlines how the field has developed over the past decade.

The quantitative analysis reveals clear asymmetries. Authorship attribution dominates the literature, with most studies addressing the identification of an author from a closed set, while only a small number consider verification. Feature extraction shows a similar pattern, the majority of papers rely solely on stylometric features derived from source code, whereas only a limited subset combines these with behavioural information. Classical machine-learning methods such as random forests and support-vector machines remain common, often used alongside neural models rather than replaced by them. Although neural architectures appear frequently, they are typically evaluated in conjunction with simpler baselines. Empirical evaluations are concentrated on a small number of datasets, most notably Google Code Jam and GitHub repositories, and reporting practices focus heavily on cross-validation and accuracy, with other metrics used less consistently. Publicly available tools, frameworks and explicit interpretability mechanisms are mentioned in only a minority of studies.

Beyond reporting frequencies, the survey synthesises relationships between features, models and datasets, and situates technical choices within broader publication and collaboration patterns. The findings indicate that programmer attribution research is sufficiently established to support systematic comparison, yet remains centred on a narrow set of benchmarks and experimental conventions. Several directions emerge from this analysis. The limited attention to verification tasks highlights the need to move beyond attribution-only settings and to study scenarios where authorship claims must be confirmed or rejected. The prevalence of stylometric features points to opportunities for incorporating behavioural evidence, such as development traces or interaction data, to capture aspects of programming that static code alone cannot reflect. Expanding dataset diversity across languages, domains and developer populations would strengthen empirical foundations, particularly by including novice programmers and educational contexts. Progress will also depend on clearer reporting of experimental design and wider adoption of evaluation practices that reflect realistic usage conditions. Finally, greater emphasis on transparency through the release of code, data and processing pipelines would support reproducibility and enable more cumulative research across the community.

\begin{acks}
This work was supported by project KEGA No.~061TUKE-4/2025 \emph{Building Bridges between University and High School ICT Education}.

ChatGPT was used to assist in the development of scripts for extracting descriptive statistics from BibTeX records and to support the design of the manuscript structure. All outputs were reviewed and validated by the authors. Grammarly was used to assist with correcting typographical and grammatical errors. The authors retain full responsibility for the content of this work.
\end{acks}

\bibliographystyle{ACM-Reference-Format}
\bibliography{sample-base}

\end{document}